\documentclass[english,aps,twocolumn,citeautoscript,superscriptaddress]{revtex4-1}
\usepackage[T1]{fontenc}
\usepackage[utf8]{inputenc}
\usepackage{babel}
\usepackage{amsmath}
\usepackage{amssymb}
\usepackage{graphicx}
\usepackage[unicode=true,pdfusetitle,
 bookmarks=true,bookmarksnumbered=false,bookmarksopen=false,
 breaklinks=false,pdfborder={0 0 1},backref=false,colorlinks=false]
 {hyperref}

\makeatletter
\date{\today}

\makeatother

\begin{document}

\title{Machine learning for molecular dynamics with strongly correlated
electrons}

\author{Hidemaro Suwa}

\affiliation{Department of Physics and Astronomy, The University of Tennessee,
Knoxville, TN 37996, USA}

\affiliation{Department of Physics, The University of Tokyo, Tokyo 113-0033, Japan}

\author{Justin S. Smith}

\affiliation{Theoretical Division and CNLS, Los Alamos National Laboratory, Los
Alamos, NM 87544}

\author{Nicholas Lubbers}

\affiliation{CCS Division, Los Alamos National Laboratory, Los Alamos, NM 87544}

\author{Cristian D. Batista}

\affiliation{Department of Physics and Astronomy, The University of Tennessee,
Knoxville, TN 37996, USA}

\affiliation{Quantum Condensed Matter Division and Shull-Wollan Center, Oak Ridge
National Laboratory, Oak Ridge, TN 37831, USA}

\author{Gia-Wei Chern}

\affiliation{Department of Physics, University of Virginia, Charlottesville, VA
22904}

\author{Kipton Barros}

\affiliation{Theoretical Division and CNLS, Los Alamos National Laboratory, Los
Alamos, NM 87544}
\begin{abstract}
We use machine learning to enable large-scale molecular dynamics (MD)
of a correlated electron model under the Gutzwiller approximation
scheme. This model exhibits a Mott transition as a function of on-site
Coulomb repulsion $U$. The repeated solution of the Gutzwiller self-consistency
equations would be prohibitively expensive for large-scale MD simulations.
We show that machine learning models of the Gutzwiller potential energy
can be remarkably accurate. The models, which are trained with $N=33$
atoms, enable highly accurate MD simulations at much larger scales
($N\gtrsim10^{3}$). We investigate the physics of the smooth Mott
crossover in the fluid phase.
\end{abstract}
\maketitle
Understanding strongly correlated electron systems is an outstanding
challenge in condensed matter theory. Some of the simplest many-body
models remain unsolved. This challenge persists despite steady progress
in theoretical and computational methods. Machine learning (ML) is
becoming a promising tool to help model various types of many-body
phenomena.

Because the many-body quantum state space grows exponentially with
system size, the cost of a direct numerical solution quickly becomes
intractable. Unbiased quantum Monte Carlo (QMC) can be very effective
in special cases (e.g., for the square-lattice Hubbard model at half
filling), but generally one is plagued by the ``sign problem'' of
resolving the delicate signal that remains after cancellations between
samples with complex phases~\cite{Troyer05}. Many clever mitigation
strategies have emerged~\cite{Parisi83,Bloch17,Cristoforetti12,Fukuma17,Frame18}.
An intriguing possibility is to use ML to extract relevant physics
from QMC samples \emph{without }reweighting~\cite{Broecker17,Chng17}.

Alternatively, one may seek to represent the many-body state variationally,
e.g. with the density matrix renormalization group ~\cite{White92,Schollwoeck11}
or tensor network generalizations. ML is inspiring new variational
ansatzes~\cite{Carleo17,Nomura17,Cai18,Levine18} that compare favorably
with previous ones.

To make quantitative predictions for real correlated electron materials,
one commonly employs physically motivated approximations, such as
variational and fixed node QMC~\cite{Foulkes01} and dynamical mean-field
theory (DMFT)~\cite{Kotliar06}. These methods remain computationally
demanding. Again, ML presents new opportunities. For example, ML may
be useful as a low-cost surrogate model for the impurity solver within
DMFT~\cite{Arsenault14} or even the full DMFT calculation itself~\cite{Arsenault15}.

An emerging research area is the molecular dynamics (MD) of strongly
correlated electron materials. Developing such toolkits not only is
of fundamental importance, but also has important technological implications.
While quantum MD methods based on density functional theory (DFT)
have been successfully applied to a wide variety of materials, they
have limited validity in their treatment of electron correlations.
On the other hand, most of the many-body techniques mentioned above
are computationally too costly for MD simulations.

ML offers the possibility of large-scale MD simulations by emulating
the time-consuming quantum calculations required at each time step.
Indeed, ML has already proven to be extremely effective in modeling
MD potentials for chemistry and materials applications~\cite{Rupp15,Behler15,Shapeev16,Smith17,Bartok17,Gilmer17,Schuett17,Lubbers18,Schuett18a,Willatt18,Dragoni18,Butler18}.
An ML model might be trained from a dataset containing $10^{4}$–$10^{6}$
individual atomic forces, often calculated with DFT. In organic chemistry,
ML now routinely predicts molecular energies that agree with new DFT
calculations to within $0.04$~eV ($1$~kcal/mol)~\cite{Smith17,Lubbers18,Schuett18a},
whereas DFT itself is almost certainly not this accurate. This success
has spurred recent efforts to calculate the training data at levels
of quantum theory significantly beyond DFT~\cite{McGibbon17,Smith18,Chmiela18}.

Here we show that ML can be similarly effective for building fast,
linear-scaling MD potentials that capture correlated electron physics.
Specifically, we use ML to enable large-scale Gutzwiller MD simulations
of a liquid Hubbard model~\cite{Chern17,Julien17}. The correlated
electronic state is computed using an efficient Gutzwiller method
at every time step. Contrary to DFT, the Gutzwiller approach captures
crucial correlation effects such as the Mott metal-insulator transition,
and produces a $U$-$T$ phase diagram qualitatively similar to that
of DMFT~\cite{Wang10}. For our neural network model~\cite{Smith17}
running on a single modern graphics processing unit (GPU), a typical
force evaluation costs $\sim10\,\mu\textrm{s}$ per atom. For the
systems considered in the present study, ML is up to six orders of
magnitude faster than direct quantum calculations. 

The present work opens a path toward large-scale dynamical simulation
of realistic models of correlated materials. Future studies could
train ML models on data generated from small-scale QMC or DMFT calculations.
ML works well assuming \emph{locality}, i.e., that the total energy
can be decomposed as a sum of local contributions~\footnote{Locality is not strictly required; long-range interactions of known
form (i.e. classical Coulomb interactions) can be added to the ML
potential by hand.}. Our success in this study offers evidence that locality can remain
valid in the presence of strong electron correlations.

Following earlier work~\cite{Chern17}, we consider a single-orbital
model Hamiltonian 
\begin{equation}
\mathcal{H}[\mathbf{r},\mathbf{p}]=\mathcal{H}_{\textrm{Hubbard}}[\mathbf{r}]+V_{\textrm{pair}}[\mathbf{r}]+E_{\textrm{kin}}[\mathbf{p}],
\end{equation}
where $\{\mathbf{r}_{i}\}$ and $\{\mathbf{p}_{i}\}$ are the positions
and momentum of the nuclei. The electronic part
\begin{align}
\mathcal{H}_{\textrm{Hubbard}} & =-\sum_{i\neq j}\sum_{\sigma}t_{ij}c_{i\sigma}^{\dagger}c_{j\sigma}+U\sum_{i}n_{i\uparrow}n_{i\downarrow},\label{eq:H_hubbard}
\end{align}
has hopping and on-site Coulomb repulsion terms. The operator $c_{i\sigma}^{\dagger}$
creates an electron with spin $\sigma\in\{\uparrow,\downarrow\}$
on the $i$th atom, and $n_{i\sigma}=c_{i\sigma}^{\dagger}c_{i\sigma}$
is the number operator. We take the system to be half filled (one
electron per nucleus). The hoppings $t_{ij}=t(|\mathbf{r}_{i}-\mathbf{r}_{j}|)$
decay exponentially with distance between nuclei, $t(r)=t_{0}\exp(-r/\xi_{1})$.
The pair repulsions $V_{\textrm{pair}}=\sum_{i\neq j}\phi_{ij}/2$
also decay exponentially, $\phi_{ij}=\phi(|\mathbf{r}_{i}-\mathbf{r}_{j}|)$
with $\phi(r)=\phi_{0}\exp(-r/\xi_{2})$. Selecting $t_{0}=24\,\textrm{eV}$,
$\xi_{1}=0.526\,\textrm{Å}$, $\phi_{0}=100\,\textrm{eV}$, and $\xi_{2}=0.283\,\textrm{Å}$
gives a highly simplified model of hydrogen. With these choices, at
$U=0$ the dimer molecule is bound with energy $-4.58\,\textrm{eV}$
at distance $r_{0}=0.83\,\textrm{Å,}$ in loose agreement with the
physical values of $-4.52\,\textrm{eV}$ and $0.74\,\textrm{Å}$.
Our model clearly departs from hydrogen, however, in the large range
of Hubbard $U$ values that we consider: $0\,\textrm{eV}\leq U\leq17\,\textrm{eV}$~\cite{Chiappe07}.
Finally, our Hamiltonian includes a kinetic energy term $E_{\textrm{kin}}=\sum_{i}|\mathbf{p}_{i}|^{2}/2m$,
where $m\approx1\,\textrm{amu}$ is the mass of the proton.

\textbf{Gutzwiller method.} To estimate the electronic free energy
at nonzero $U$, we employ a finite temperature generalization of
the Gutzwiller projection method~\cite{Wang10,Sandri13}. In this
approach, we seek a variational approximation $\rho_{G}=\mathcal{P}\rho_{0}\mathcal{P}$
for the density matrix $\exp(-\beta\mathcal{H}_{\textrm{Hubbard}})$.
Here $\rho_{0}$ is the Boltzmann distribution of free quasi-particles,
and the so-called Gutzwiller projection operator $\mathcal{P}=\prod_{i}\mathcal{P}_{i}$
effectively reweights electron occupation numbers at each site. We
consider the paramagnetic phase of the Hubbard model, which means
$n_{i,\uparrow}=n_{i,\downarrow}$. The variational target is to minimize
a free energy $F_{G}=\langle\mathcal{H}_{\textrm{Hubbard}}\rangle_{G}-TS_{G}$,
where $S_{G}=\textrm{Tr}\,(\rho_{G}/Z_{G})\ln(\rho_{G}/Z_{G})$, subject
to the Gutzwiller constraint $\langle n_{i}\rangle_{G}=\langle n_{i}\rangle_{0}$.
We use $\langle\mathcal{O}\rangle_{G}$ and $\langle\mathcal{O}\rangle_{0}$
to denote expectations of $\mathcal{O}$ computed from density matrices
$\rho_{G}$ and $\rho_{0}$, respectively. Importantly, the expectation
values of the hopping terms can be efficiently computed using the
so-called Gutzwiller approximation~\cite{Gutzwiller65} (exact in
infinite dimensions): $\langle c_{i}^{\dagger}c_{j}\rangle_{G}=\mathcal{R}_{i}\mathcal{R}_{j}\,\langle c_{i}^{\dagger}c_{j}\rangle_{0}$,
where the renormalization factor $\mathcal{R}_{i}$ is uniquely determined
from the electron density $n_{i}^{0}=\langle n_{i}\rangle_{G}$ and
double occupancy $d_{i}=\langle n_{i\uparrow}n_{i\downarrow}\rangle_{G}$~\cite{Lanata12}. 

Although $S_{G}$ cannot be evaluated exactly, a good approximation
to the free energy is $F_{G}=F_{0}+U\sum_{i}d_{i}-T\Delta S.$ The
term $F_{0}=-k_{B}T\ln{\rm Tr}\rho_{0}$ is the free energy of quasi-particles
and $\Delta S$ is the entropy correction due to the projector $\mathcal{P}$~\cite{Wang10}.

Self-consistent minimization of $F_{G}$ with respect to the variational
parameters produces the electronic free energy of interest. This minimization
is performed by cycling between two steps~\cite{Lanata12}. (1) For
fixed $n_{i}^{0}$ and $d_{i}$, the renormalized Hamiltonian
\begin{equation}
\mathcal{H}_{0}=-\sum_{i\neq j}\mathcal{R}_{i}\mathcal{R}_{j}t_{ij}c_{i}^{\dagger}c_{j}-\sum_{i}\mu_{i}n_{i}\label{eq:H_0}
\end{equation}
is diagonalized to obtain $\rho_{0}$. The Lagrange multipliers $\mu_{i}$
impose the density constraint. (2) For fixed $\rho_{0}$, one adjusts
$n_{i}^{0}$ and $d_{i}$ to minimize $F_{G}$.

Once converged, we treat $V_{\textrm{elec}}=\min\,F_{G}$ as the electronic
part of the total MD potential, $V=V_{\textrm{elec}}+V_{\textrm{pair}}$.
The corresponding forces, 
\begin{equation}
-\frac{\partial V}{\partial\mathbf{r}_{i}}=2\sum_{j}\frac{\partial t_{ij}}{\partial\mathbf{r}_{i}}\mathcal{R}_{i}\mathcal{R}_{j}\,\langle c_{i}^{\dagger}c_{j}\rangle_{0}-\sum_{j}\frac{\partial\phi_{ij}}{\partial\mathbf{r}_{i}},\label{eq:force}
\end{equation}
drive MD simulations of the nuclei under the Born-Oppenheimer approximation.
To derive Eq.~(\ref{eq:force}), we used the fact that $F_{G}$ is
minimized with respect to the variational parameters.

The scheme here is largely similar to that of our previous Gutzwiller
MD study~\cite{Chern17}. The primary difference is that, here, we
\emph{reinitialize} the variational parameters at each MD time step
before iteratively optimizing them. In the prior version of our code,
we selected the initial guess for ($n_{i}^{0}$, $d_{i}$) as the
self-consistent solution obtained in the previous MD time step. That
lack of reinitialization leads to weakly stable solution branches
of the Gutzwiller equations, and strong hysteresis. In the present
study, we enforce that $V_{\textrm{elec}}$ is single valued by reinitializing
$(n_{i}^{0},d_{i})$ to default values at each time step before iteratively
solving the self-consistency equations. This scheme eliminates hysteresis
and simultaneously lowers the Gutzwiller variational free energy.

\textbf{Machine learning.} Solving the above Gutzwiller equations
for the MD potential $V_{\textrm{elec}}[\mathbf{r}]$ at each time
step can be computationally expensive. We use an ML model to estimate
$\hat{V}_{\textrm{elec}}\approx V_{\textrm{elec}}$, while treating
$V_{\textrm{pair}}$ exactly. A key assumption is the (non-unique)
decomposition of energy as a sum of local contributions, $\hat{V}_{\textrm{elec}}=\sum_{i=1}^{N}\hat{V}_{\textrm{elec};i},$
where $V_{\textrm{elec};i}$ is a function only of the atomic environment
near atom $i$. We use a neural network to model the local potential
$V_{\textrm{elec};i}$. We have explored two established architectures,
the hierarchically interacting particle neural network (HIP-NN)~\cite{Lubbers18}
and the accurate neural network engine for molecular energies (ANI)~\cite{Smith17}.
Although HIP-NN may yield slightly better accuracies, we selected
ANI for our MD simulations because of its highly optimized NeuroChem
implementation~\footnote{ANI NeuroChem implementation, \url{https://github.com/isayev/ASE_ANI},
{[}Accessed: 1-January-2019{]}}.

ANI constructs a rotationally and translationally invariant representation
of the environment near atom $i$ from two types of information: (1)
the pairwise distances $\{r_{ij}\}$ for all atoms $j$ satisfying
$r_{ij}=|\mathbf{r}_{i}-\mathbf{r}_{j}|<2.8\,\textrm{Å}$, and (2)
the set of three-body angles $\{\theta_{jik}\}$, where $\cos\theta_{jik}=\mathbf{r}_{ij}\cdot\mathbf{r}_{ik}/r_{ij}r_{ik}$,
provided that both $r_{ij}$ and $r_{ik}$ are less than $2.0\,\textrm{Å}$.
The values $\{r_{ij}\}$ and $\{\theta_{jik}\}$ transformed into
a fixed-length ``feature vector'' $\{G_{i,m}\}_{m=1\dots M_{0}}$
using continuous binning. In this study, $G_{i,m}$ contains $M_{0}=96$
scalar components. There will typically be about 5–15 atoms within
a distance $2.8\,\textrm{Å}$ of atom $i$, and $G_{i,m}$ describes
this environment.

\begin{figure}
\includegraphics[width=0.95\columnwidth]{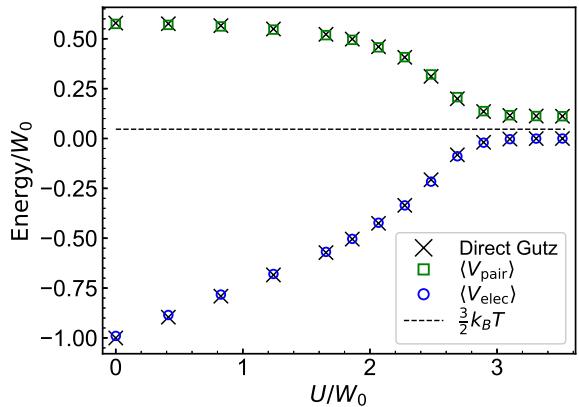}\caption{\label{fig:energies}MD averaged electronic free energy $\langle V_{\textrm{elec}}\rangle$
and pair repulsion energy $\langle V_{\textrm{pair}}\rangle$, for
varying Hubbard $U$. At every MD time step, forces are produced by
an ML model that emulates the Gutzwiller calculation. The crosses
represent reference simulations that use direct Gutzwiller calculations
rather than ML. The dashed line represents the kinetic energy. The
energy scale is $W_{0}=-\langle V_{\textrm{elec}}\rangle|_{U=0}=4.84\,\textrm{eV}$.
Tests show that these results with $N=100$ atoms are representative
of $N\rightarrow\infty$.}
\end{figure}

The neural network $\hat{V}_{\textrm{elec},i}$ is composed of multiple
real-valued activations $\{z_{m}^{\ell}\}_{m=1\dots M_{\ell}}$ for
layer index $\ell=0\dots L$. The input to the neural network is the
feature vector, $z_{m}^{\ell=0}=G_{i,m}$. Each neural network layer
has the form, $z_{m}^{\ell+1}=f_{\textrm{activ}}(\sum_{n=1}^{M_{\ell}}w_{mn}^{\ell}z_{n}^{\ell}+b_{m}^{\ell})$.
The matrix elements $w_{mn}^{\ell}$ and offsets $b_{m}^{\ell}$ are
learnable parameters. We select $f_{\textrm{activ}}(x)$ to be the
continuously differentiable exponential linear unit (CELU)~\cite{Barron17}.
A linear combination of the  outputs of the final layer yields the
local potential, $\hat{V}_{\textrm{elec};i}=\sum_{n=1}^{M_{L}}w_{n}^{L}z_{n}^{L}+b^{L}$.
We select $L=3$ and layer sizes $M_{0}=96$, $M_{1}=48$, $M_{2}=32$,
and $M_{3}=16$. There are thus approximately $10^{4}$ learnable
parameters in this neural network.

Training of the model parameters involves optimizing a loss function
$\mathcal{L}$ that quantifies the deviation between the model $\hat{V}_{\textrm{elec}}$
and direct Gutzwiller calculations $V_{\textrm{elec}}$, evaluated
on a training dataset. The loss function incorporates errors in the
potential, $\hat{V}_{\textrm{elec}}-V_{\textrm{elec}}$ and forces
$\nabla_{\mathbf{r}_{i}}(\hat{V}_{\textrm{elec}}-V_{\textrm{elec}})$.
We optimize the parameters $w_{mn}^{\ell}$ and $b_{m}^{\ell}$ using
the adaptive moment estimation (Adam) variant of stochastic gradient
descent~\cite{Kingma14}. To mitigate overfitting, we employ weight
decay regularization with $\alpha=10^{-6}$~\cite{Loshchilov17}
and early stopping. For a general introduction to this ML methodology,
see Ref.~\onlinecite{Goodfellow16}.

\begin{figure}
\includegraphics[width=0.95\columnwidth]{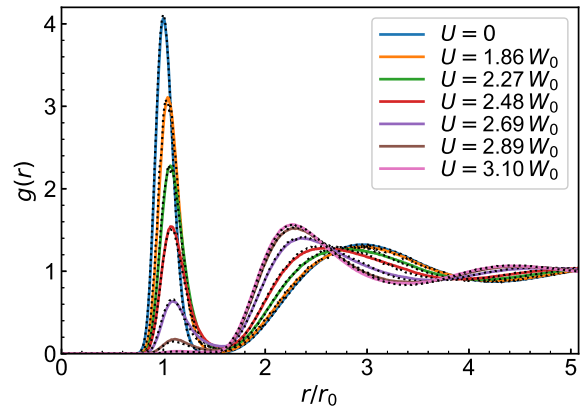}\caption{\label{fig:rdf}Radial distribution functions $g(r)$ for varying
Hubbard $U$, and the same simulation parameters as in Fig.~\ref{fig:energies}.
The results using ML (colored lines) are in strong agreement with
direct Gutzwiller calculations (black dots). The typical bond length
at $U=0$ is $r_{0}=0.83\,\textrm{Å}$.}
\end{figure}

To produce our training dataset, we ran MD simulations with forces
obtained from direct Gutzwiller calculations. We used Verlet integration
with a time step of 0.2~fs to evolve the atoms. The simulation box
has periodic boundary conditions, and its volume $V_{0}$ is set according
to a fixed density $\varrho_{0}=N/V_{0}=0.153\,\textrm{Å}^{-3}$.
Just $N=33$ atoms are sufficient to train an extensible ML potential,
which remains valid for much larger $N$. To fix the temperature $k_{B}T=0.15\,\textrm{eV}$,
we employ a Langevin thermostat with a friction coefficient $\gamma=5\times10^{-3}\,{\rm amu/fs}$
. We generated independent training data sets for $U$ values ranging
from 0 to 17~eV. For each $U$, we collected $3000$ snapshots, one
per $200$ MD time steps. Every snapshot contains the electron free
energy and associated forces.

We found that an ML model trained on this dataset alone would lead
to unstable MD simulations. To improve the robustness of our ML model,
we collected additional data that sampled a broader range of the atomic
configurations. Specifically, we augmented our training dataset by
running additional direct Gutzwiller MD with $k_{B}T=0.075\,\textrm{eV}$
and $k_{B}T=0.3\,\textrm{eV}$ for the Langevin thermostat only, \emph{without}
changing the temperature used in the Gutzwiller equations. Our total
training dataset, per $U$ value, thus contains $9,000$ MD snapshots
and about $3\times10^{5}$ atomic forces.

We reduce variance by averaging over an ensemble of eight neural networks,
each trained on a subset of the data~\cite{Smith18a}.

\begin{figure}
\includegraphics[width=0.95\columnwidth]{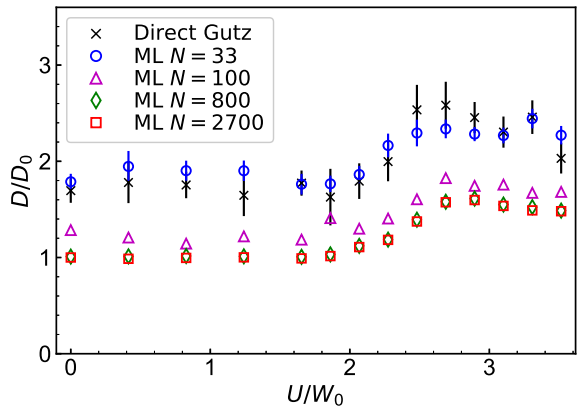}\caption{\label{fig:diffusion}The diffusivity $D$ for varying Hubbard $U$,
and the same simulation parameters as in Fig.~\ref{fig:energies}.
At $U=0$ with $N=2700$ atoms, we measure $D_{0}=0.0088\,\textrm{Å}^{2}/\textrm{fs}$.
Direct Gutzwiller calculations (black crosses) employ five times fewer
time steps than used in the ML-based simulations. Approximately $N=800$
atoms are required to reach convergence.}
\end{figure}

\textbf{Results.} We use our ML models to drive efficient and accurate
MD simulations at large scales. Figure~\ref{fig:energies} shows
mean energies for an MD simulation with $N=100$ atoms (keeping $k_{{\rm B}}T=0.15\,\textrm{eV}$)
and $3\times10^{6}$ time steps. The mean electronic energy evaluated
at $U=0$ serves as a convenient energy scale, $W_{0}=4.84\,\textrm{eV}$.
At large $U$ the electrons become localized, the electronic energy
$V_{\textrm{elec}}$ goes to zero, and atoms repel according to $V_{\textrm{pair}}$.
At small $U$ the system is metallic and $V_{\textrm{elec}}$ is an
attractive interaction that tends to bond atoms into dimer molecules,
counterbalanced by $V_{\textrm{pair}}$.

For validation, we also performed reference simulations using forces
from direct Gutzwiller calculations at each MD time step, and only
$\approx10^{5}$ time steps. The ML and reference simulations in Fig.~\ref{fig:energies}
are barely distinguishable, which is remarkable given that the training
data were generated using only $N=33$ atoms. We also directly compare
the ML predictions $\hat{V}_{\textrm{elec}}$ with reference energy
calculations $V_{\textrm{elec}}$ for $N=100$ and $N=800$. The mean
absolute error (MAE) scales as $0.02\,\sqrt{N}\,W_{0}$. The factor
$\sqrt{N}$ appears because the errors in local contributions $\hat{V}_{\textrm{elec};i}$
and $\hat{V}_{\textrm{elec};j}$ are essentially independent for atoms
$i\neq j$. The MAE of electronic force $-\nabla\hat{V}_{\textrm{elec}}$
is approximately $0.009\,f_{0}$, where $f_{0}=17.7\,\textrm{eV}/\textrm{Å}$
is the mean force magnitude at $U=0.$ If we were to reduce our training
data by a factor of 10, this force MAE would increase by about 10\%.

Figure~\ref{fig:rdf} shows the radial distribution functions $g(r)$.
At $U=0$, $g(r)$ has a characteristic peak at $r_{0}=0.83\,\mathring{A}$,
which reflects the bond length of a dimer molecule. This peak gradually
decreases with increasing $U$, and disappears entirely at $U\gtrsim3.1\,W_{0}$,
where $\langle V_{\textrm{elec}}\rangle\approx0$ according to Fig.~\ref{fig:energies}.
The ML-based simulations again appear almost identical to direct-Gutzwiller
reference simulations.

Figure~\ref{fig:diffusion} shows a \emph{dynamic} observable, the
diffusivity
\begin{equation}
D=\frac{1}{3N}\sum_{i=1}^{N}\int_{0}^{\infty}\langle{\bf v}_{i}(t)\cdot{\bf v}_{i}(0)\rangle\mathrm{d}t.\label{eq:D}
\end{equation}
Replacing $\infty$ with a finite time $\tau$ would yield a naive
estimator of $D$. Instead we use an estimator $\hat{D}(\tau)=\frac{1}{2\tau}\frac{1}{3N}\sum_{i}|\int_{0}^{\tau}{\bf v}_{i}(t)dt|^{2}$
with reduced variance. Varying $\tau$, we find that the bias of the
estimator $D\approx\langle\hat{D}(\tau_{0})\rangle$ becomes negligible
at $\tau_{0}$ corresponding to $10^{5}$ MD time steps. We collected
more than ten independent samples of $\hat{D}(\tau_{0})$ from each
MD simulation and estimated the error bars using bootstrapping. $D$
converges surprisingly slowly with system size; about $N\approx800$
atoms seem to be required. Such simulations would have been extremely
challenging without ML.

Due to updates in our Gutzwiller solver, the present results deviate
significantly from prior work~\cite{Chern17}. The mean energies
(Fig.~\ref{fig:energies}) and $g(r)$ curves (Fig.~\ref{fig:rdf})
now vary smoothly with $U$, indicating a crossover rather than a
first-order transition. We also observe in Fig.~\ref{fig:diffusion}
that $D$ decays smoothly after reaching its peak at the Mott transition,
whereas before we had observed a sharp drop. The previous results
were dependent on strong hysteresis effects. Here we reinitialize
the Gutzwiller parameters at every time step, which eliminates hysteresis
and lowers the variational free energy in all instances we checked.
We argue that the present approach is more consistent with the assumption
(used in the finite-temperature Gutzwiller method) that the electrons
are in equilibrium.

\begin{figure}
\includegraphics[width=0.95\columnwidth]{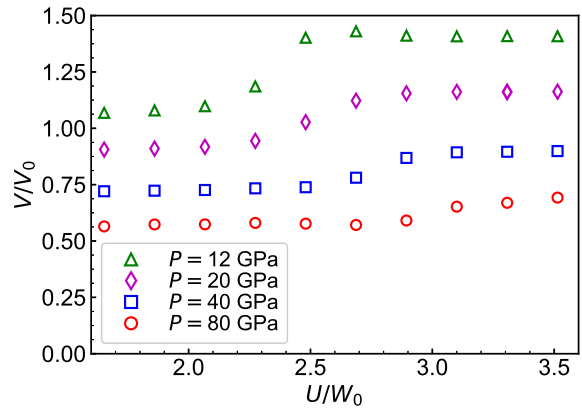}\caption{\label{fig:density}A volume collapse occurs with decreasing $U$.
Here we employ an ensemble with constant pressure $P=12,20,40,$ and
$80\,\textrm{GPa}$; the pressure of the previous simulations at fixed
volume $V_{0}(N)=N/\varrho_{0}$ corresponds to $P=20\,\textrm{GPa}$
when $U\approx2.4\,W_{0}$. At this temperature, there is a relatively
smooth Mott cross-over. }
\end{figure}

Finally, we investigate the nature of the Mott transition using large-scale
MD with $N=2700$ atoms. Here, we switch to an ensemble with constant
pressure $P=12,20,40,$ and $80\,\textrm{GPa}$, implemented using
the Monte Carlo barostat algorithm~\cite{Faller02}. The pressure
of our previous (fixed density) simulations matches $P=20\,\textrm{GPa}$
when $U\approx2.4\,W_{0}$. Figure~\ref{fig:density} shows collapse
of volume $V$ with decreasing $U$, corresponding to the crossover
from the Mott insulating phase to the metallic state. 

The ML techniques applied here demonstrate that, with more realistic
quantum models, it will be possible to perform large-scale Gutzwiller
MD simulations of the structural properties of real \emph{f}-electron
materials, such as Pu, and other correlated electron systems. It is
surprising how accurately ML predicts Gutzwiller forces using only
the atomic environment within a 2.8~Å radius. The power of ML is
that, under the locality assumption, only training data for relatively
small system sizes are required. In future work, it should be feasible
to generate training data from, e.g., quantum Monte Carlo or DMFT.
Once the ML model has been trained, accurate MD simulations of unprecedented
scale become practical. For MD boxes with $N=2700$ atoms, our ML
model running on a Tesla P100 GPU can evaluate all forces in about
26~ms, which is $\approx10^{6}$ times faster (extrapolated) than
our reference Gutzwiller implementation running on a Intel Xeon CPU
E5-2680. We anticipate that a distributed ML/MD implementation could
readily enable simulations of $f$-electron materials with millions
of atoms.

\acknowledgments

H. Suwa, C. D. Batista, and G.-W. Chern acknowledge support from the
center of Materials Theory as a part of the Computational Materials
Science (CMS) program, funded by the U.S. Department of Energy, Office
of Science, Basic Energy Sciences, Materials Sciences and Engineering
Division. J. S. Smith, N. Lubbers, and K. Barros acknowledge support
from the LDRD and ASC PEM programs at LANL. Computer simulations were
performed using IC and Darwin resources at LANL.

\bibliographystyle{apsrev4-1}
\bibliography{gutz_ml}

\end{document}